\documentclass[12pt,nofootinbib]{article}

\usepackage{amssymb,graphicx,slashed} 
\usepackage{epsf}
\usepackage{pstricks}

\newcommand{\nc}{\newcommand}
\nc{\postscript}[2]{\setlength{\epsfxsize}{#2\hsize}\centerline{\epsfbox{#1}}}

\def\beq{\begin{equation}}
\def\eeq{\end{equation}}
\def\bea{\begin{eqnarray}}
\def\eea{\end{eqnarray}}

\def\bit{\begin{itemize}}
\def\eit{\end{itemize}}

\def\l{\left}
\def\r{\right}

\def\ra{\rightarrow}
\def\baa{\begin{array}}
\def\eaa{\end{array}}
\def\Qc{{\cal Q}}
\def\Bc{{\cal B}}
\def\Tc{{\cal T}}

\def\d{\partial}

\def\simgt{\mathrel{\lower2.5pt\vbox{\lineskip=0pt\baselineskip=0pt
           \hbox{$>$}\hbox{$\sim$}}}}
\def\simlt{\mathrel{\lower2.5pt\vbox{\lineskip=0pt\baselineskip=0pt
           \hbox{$<$}\hbox{$\sim$}}}}
\newcommand{\vev}[1]{ \langle {#1} \rangle }
\begin{document}
\begin{titlepage}
\begin{flushright}
\end{flushright}

\vskip.5cm

\begin{center}
{\huge \bf  Light Custodians and Higgs Physics}
\vskip.3cm
{\huge \bf  in Composite Models} 
\end{center}
\vskip1cm

\begin{center}
{\bf Aleksandr Azatov and Jamison Galloway}
\end{center}

\begin{center}
{\it Dipartimento di Fisica, Universit\`a di Roma ``La Sapienza'' \\
and INFN Sezione di Roma, I-00185 Roma, Italy} \\
\vspace*{0.3cm}
{\tt  aleksandr.azatov@roma1.infn.it, jamison.galloway@roma1.infn.it}
\end{center}

\vglue 0.3truecm

\begin{abstract}
\vskip 3pt \noindent
Composite Higgs models involving partial compositeness of Standard Model fermions  typically require the introduction of fermionic partners which are relatively light in realistic scenarios. 
In this paper, we analyze the role  of these light custodian fermions in the phenomenology of the composite Higgs models and show that they  significantly modify couplings of the Higgs field.  
We focus on the coupling to gluons in particular which is of central importance for Higgs production at the LHC. We show that this coupling can be increased as well as decreased depending on the Standard Model fermion embedding in the composite multiplets. We also discuss modification of the Higgs couplings to bottom and top quarks and show that modifications to all  three couplings---$Hgg$, $H \bar tt$, and $H \bar bb$---are generically independent parameters.
\end{abstract}

\end{titlepage}



\section{Introduction}
\label{sec:Intro} \setcounter{equation}{0} \setcounter{footnote}{0}
Composite Higgs models provide a compelling solution to the Standard Model (SM) hierarchy problem \cite{Kaplan:1983sm, Kaplan:1983fs,Georgi:1984af,Dugan:1984hq}. Within these models, the stability of the Higgs mass against large quantum corrections is explained by the strongly interacting nature of the Higgs field.  From the precision measurements performed at LEP2  \cite{Flacher:2008zq}, we in fact have the strong suggestion that the mass of the Higgs should be less than roughly 170 GeV.  Such a light resonance from the strong sector can be {\it naturally} obtained if one assumes that the Higgs emerges as a pseudo Nambu-Goldstone boson (PNGB) of a spontaneously broken global symmetry, making it much lighter than the other, non-Goldstone, resonances \cite{Kaplan:1983sm, Kaplan:1983fs,Georgi:1984af,Dugan:1984hq,Giudice:2007fh}. 

For fermions to acquire mass in these models, one can further suppose that the strong sector of the model contains composite fermions with which the SM states can mix: this is the idea of partial compositeness \cite{Kaplan:1991dc}.   In this way, the hierarchies of the SM fermion masses   are explained by the hierarchies of the correponding mixing parameters, as can also be easily realized in Randall-Sundrum models \cite{Randall:1999ee}
 with all the SM fields in the bulk. PNGB nature of the Higgs boson can be realized in 5D framework
\cite{
Contino:2003ve,Agashe:2004rs}, where the Higgs field arises as an extradimenional component of the 5D gauge field.
In such models it often happens that most  of the new resonances are too heavy to be discovered early in the running of the LHC,   
but indirect effects resulting in the observable modifications of the  SM Higgs couplings  can be quite important.
In these scenarios, it is crucial to understand the predictions for deviations from various SM couplings. 

The dominant production channel of the Higgs field at LHC is the process of gluon fusion, as is generated in the SM by a loop of top quarks.   In the composite Higgs models, this coupling will be modified by two separate effects: first, one generically finds a modification of the top Yukawa coupling, and second one must account for loops of the additional new states. These effects were studied previously for the composite models, both for cases where the Nambu-Goldstone mechanism was invoked \cite{Falkowski:2007hz,Low:2009di,Low:2010mr,Furlan:2011uq}, and for those where it was not \cite{Lillie:2005pt,Djouadi:2007fm,
Bouchart:2009vq,Casagrande:2010si,Azatov:2010pf}. In this paper we will analyze the structure of the $Hgg$ coupling within composite PNGB Higgs models. First we will review the previous analyses of the Higgs couplings in the PNGB models, where in all cases a suppression of the $Hgg$ coupling  has been reported.
 We will show that this is not the case in general, however,
and identify conditions a model must satisfy to lead to an {\it enhancement} of this interaction, and construct example models.  
We show that the enhancement of the $Hgg$ coupling can happen  due to the effects of the composite partners of a $b$ quark in the models where $b_L$ is fully composite, or due to the effects of composite partners of the top quark, in the models where the SM top quark mass is generated by more than one operator. We also demonstrate that the Higgs couplings ($Hgg$, $H\bar tt$, $H\bar bb$) are generically modified in an independent way.

The outline of the discussion is as follows: In Section~\ref{sec:Review}, we  present the composite Higgs model based on  the minimal coset $SO(5)/SO(4)$.  In Section~3, we review the effects on the $Hgg$ coupling from arbitrary heavy fermion fields. In Section~\ref{toyexample}, we then show two toy examples based on composite models with a single composite fermion multiplet (considering separately the case of the ${\bf 5}$ and the ${\bf 10}$),  illustrating the effects of the light custodian $t'$ and $b'$ fields on the SM Higgs couplings $Hgg$, $H\bar tt$, and $H\bar bb$.
In Section~5, we finally discuss a realistic 5D-inspired  composite Higgs model.
We  discuss the Higgs couplings in the model and  show the numerical results for the modifications of the promising light Higgs signal $gg\ra H \ra \gamma\gamma$.   We  discuss  also the bounds
on the model from LEP,  the Tevatron, and recently from the LHC.  We conclude in Section~6.

\section{Partial Compositeness and a PNGB Higgs}
\label{sec:Review} \setcounter{equation}{0} \setcounter{footnote}{0}
In this section we will highlight some basic features of models with partially composite fermions as well as a composite Higgs; it is in these cases that we find potentially significant deviations compared to the SM.  We take as our basic setup that of a two-site model, where states are classified as either purely composite or purely elementary.  Soft mass mixing of the two sectors is then introduced in order to describe {\it partially} composite states.  A thorough discussion of this scenario and its central  philosophy can be found in \cite{Contino:2006nn}(see also \cite{DeCurtis:2011yx}); here we simply recall its interpretation as a simplified picture of warped five-dimensional setups in which  the degree of compositeness of a given state is determined by its  localization along the extra dimension.  The two-site model is regarded as a deconstructed version of such a picture, where the two sites correspond loosely to the endpoints of an extra dimension.
 We turn now to a brief review of how fermions and scalars are described in the two-site language.

\subsection{The Scalar Sector}
The scalar sector of the model consists of the Higgs field and the longitudinal components of the $W$ and $Z$.  In our setup, these states arise as Nambu-Goldstone bosons (`pions') of a spontaneously broken global symmetry, with the Higgs acquiring its mass from some small explicit breaking of the global symmetry.  It is because of its role as a  PNGB that the Higgs can be naturally light.  As such, the scalars are conveniently expressed as fluctuations on the coset manifold about some vacuum orientation.  In general, we can encode them in a matrix $\xi(x)$, such that a fundamental representation of the global group is constructed as in the Callan-Coleman-Wess-Zumino (CCWZ) prescription \cite{Coleman:1969sm,Callan:1969sn}:
\bea
\Sigma(x)  &=& \xi(x) \cdot \Sigma_0 \nonumber \\
&=& \exp(i \Pi/f) \cdot \Sigma_0.
\eea
Here $\Pi$ is a sum of pion fields along each broken direction, and $\Sigma_0$ corresponds to a chosen vacuum orientation, which we will refer to  as the `standard' vacuum.   Recall that the pions themselves transform nonlinearly, $\xi(x) \mapsto {\cal G}\cdot \xi(x) \cdot {\cal H}^{-1}$, with ${\cal G}$ an element of the global symmetry and ${\cal H}$ an element of the subgroup left unbroken by the spontaneous breaking.

We will focus in this paper on the specific coset $SO(5)/SO(4)$ as examined in \cite{ Agashe:2004rs}, though our results are straightforward to apply to models based on larger spaces, e.g. those in \cite{Gripaios:2009pe,Mrazek:2011iu}.  In addition, the composite sector contains a global $U(1)_X$, whose presence is required in order to correctly reproduce electric charges for the fermion fields as we will see below.  The symmetry breaking pattern we work with is chosen simply because it is the minimal coset that provides the three longitudinal gauge components as well as a light Higgs state, the latter of which is strongly suggested to exist by precision data.  

With the group generators specified in Eq.~(\ref{eq:generators}), the standard vacuum takes the form
\beq\label{eq:Sigma0}
\Sigma_0 = \begin{pmatrix} {0 & 0 & 0 & 0 & 1} \end{pmatrix}^T.
\eeq
We assume that the dynamics is such that the physical vacuum acquires an electroweak symmetry breaking (EWSB) component once the Higgs satisfies $\vev {H(x)}\neq 0$ .   That is,
\beq\label{eq:vacuum}
\Sigma (x) = \begin{pmatrix} 
{ 0 \cr 0 \cr 0 \cr \sin \frac{H(x)}{f} \cr \cos \frac{H(x)}{f} }
\end{pmatrix},
\eeq
which can be understood as a rotation of the standard vacuum, taking
\bea\label{eq:xi}
\xi(x) &=&  \exp(i \sqrt 2 \, T_C^4 \, H(x)/f) \nonumber \\
&=& \begin{pmatrix} 
{{\bf 1}_3 & 0 & 0 \cr
0 & \cos \frac{H(x)}{f}  & \sin \frac{H(x)}{f} \cr
0 & -\sin \frac{H(x)}{f} & \cos \frac{H(x)}{f} }
\end{pmatrix}.
\eea

\subsection{The Fermion Sector}
In the two-site setup, SM quarks and leptons acquire their mass by mixing with composite states, the latter of which are assumed to be the only fermions that interact directly with the PNGB Higgs.  Here we will focus on two possible representations in which fermions can be introduced, namely the $\bf 5$ and the $\bf 10$.  This allows the following  $SO(5)$-invariant interactions at quadratic order in the fermion fields:
\bea
\label{globalint}
\Delta {\cal L} =  (\bar {\bf 5} \cdot \Sigma)^2+(\Sigma^\dagger \cdot \bar{{\bf 10}} \cdot {\bf 10}  \cdot \Sigma)+ (\bar {\bf 5} \cdot {\bf 10} \cdot \Sigma)+{\rm h.c.}
\eea
Note that using a simple fermion field redefinition,  the Higgs can be made to  appear  only in the derivative couplings (see discussion in section \ref{toyexample}), such that the Goldstone symmetry of $h(x)$ is respected.

To begin, we consider an example where a generation  of composite `quarks'---containing partners for the top and bottom quarks---is  embedded in a ${\bf 5}$ with $U(1)_X$ charge $2/3$.  Using the decomposition as described in the appendix, the composite has the $SO(5)$ form
\beq\label{eq:Q5}
\Qc = \frac{1}{\sqrt 2} \begin{pmatrix}
{\chi + B  \cr
i(\chi-B) \cr
T+T' \cr
i(T-T')\cr
\sqrt 2 \tilde T}
\end{pmatrix},
\eeq
decomposing under $SU(2)_L \times SU(2)_R$ as a bidoublet and singlet
\beq
\begin{pmatrix} {T & \chi \cr
B & T'}
\end{pmatrix} \oplus \tilde T.
\eeq
The $T$'s all have electric charge $2/3$, $B$ has charge $-1/3$, and the exotic $\chi$ has charge $5/3$ (see appendix \ref{generators} for details).  

To demonstrate how the  SM fermions acquire their mass, we review a simple example with matter content summarized in Table~\ref{1GenMatter}. 
\begin{table}[ht]
\begin{center}
\begin{tabular}{c|c c}
     &  $SU(2)_L$  & $U(1)_Y$  \\ \hline
$q_L$ & $\square$ & 1/6 \\
$t_R$ & 1 & 2/3 \\
$ b_R$ & 1 & -1/3 \\
$Q = P_q \Qc $ & $\square$ & 1/6 \\
$T = P_t \Qc $ & $1$ & 2/3 
\end{tabular}\caption{Fermionic content of the two-site model needed to describe the third generation of quarks.  The bottom quark will remain massless, as there is no mixing that can involve $b_R$.}\label{1GenMatter}
\end{center}
\end{table} 
The composite sector is assumed to contain a single vector-like $\bf 5$, Eq.~(\ref{eq:Q5}).  This implies that in order to write electroweak gauge-invariant interactions, we must first project out the components of this multiplet with the quantum numbers of the left- and right-handed top quark.  These projectors are denoted respectively $P_q$ and $P_t$, with explicit forms that can be determined by comparison with Eqs.~(\ref{eq:Q5}) and (\ref{eq:generators}).    We will use this notation throughout: script capital letters denote the $SO(5)$ composite multiplets, while plain capital letters denote the projections onto EW gauge  multiplets.

Now we  include mass terms for the composites, and soft mixing terms with the elementary fields:
\bea\label{eq:SimpleMasses}
\Delta {\cal L}= M_{\Qc}\bar \Qc \Qc + \left( \lambda_q \bar q_L Q_R+\lambda_t \bar t_R  T_L +{\rm h.c.}\right).
\eea
With this, rotating by an angle $\theta$ defined by 
\beq
\tan \theta_q=\frac{\lambda_q}{M_{\Qc}}
\eeq
we find that before EWSB we have the following massless states:
\bea\label{eq:SMstates}
q^{\rm SM}_L=q_L \cos \theta  - Q_L \sin \theta,
\eea
and the orthogonal combination
\bea
\tilde Q = Q_L \cos \theta + q_L \sin \theta_q,
\eea
with mass
\beq
M_{\tilde Q}=\frac{\lambda_q}{\sin \theta_q}.
\eeq
From Eq.~(\ref{eq:SMstates}), we see that SM fermion mass hierarchies will arise from hierarchies of the couplings $\lambda_i$.  Heavy fields are understood to have a larger degree of compositeness, giving stronger interactions with the composite Higgs.

Note also that the mass of the fields in the global multiplet $\Qc$ that do not mix  elementary fields will be given simply by $M_{\Qc}$,
which tends to zero in the limit $\sin \theta_q\ra 1$ with $\lambda$ fixed,  corresponding  to full compositeness of the SM field. This behavior is as  expected from holographic theories, where custodian fields are becoming light in such a limit.

Finally, we note that the couplings $\lambda_i$  break the Goldstone symmetry of the Higgs boson, leading to the appearance of nonderivative interactions at higher orders.  This will be crucial when we come to calculating the contribution from light custodians to Higgs couplings, as the interactions of interest themselves explicitly violate the Goldstone symmetry.  This allows one to treat the couplings $\lambda_i$ as spurions in order to understand the structure of loop-induced interactions.  This will be seen explicitly below.

\section{$Hgg$ from Integrating out Heavy Fermions}
\label{sec:Intro} \setcounter{equation}{0} \setcounter{footnote}{0}
\begin{figure}
\begin{center}
\includegraphics{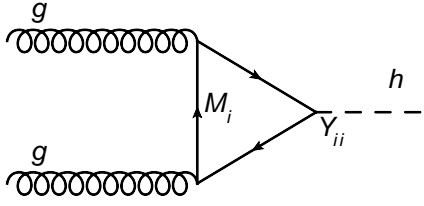}
\caption{Higgs coupling to gluons induced by a loop of massive fermions. \label {hggloop}}
\end{center}
\end{figure}

Before proceeding to the calculation of the $Hgg$ coupling in the composite models let us review the calculation of this coupling in  a generic model with heavy fermions. 
The calculation simplifies in the physical mass basis (as defined after the Higgs develops a VEV), i.e. where
\bea
\Delta {\cal L}=\sum M_i(v)\bar\psi_i\psi_i+\sum Y_{ij}  
\bar\psi_i\psi_j H(x).
\eea 
The fermion contribution to the Higgs production cross-section from gluon fusion is given by \cite{Ellis:1975ap}
\bea
\sigma_{gg \to H}^{SM} &=& \frac{\alpha_s^2 m_H^2}{576 \pi }\left|\sum_i \frac{Y_{ii}}{M_i} A_{1/2}(\tau_i)\right|^2 \delta(\hat{s}-m_H^2), 
\eea
where
\bea
\tau_i &\equiv& m_h^2/4M_i^2, \nonumber\\
A_{1/2}(\tau) &=& \frac{3}{2}[\tau + (\tau-1)f(\tau)]\tau^{-2},  \\
f(\tau) &=&\l \{\baa{c}[\arcsin\sqrt{\tau}]^2,\quad(\tau \le 1), \nonumber  \\
-\frac{1}{4}\left[\ln\left(\frac{1+\sqrt{1-\tau^{-1}}}{1-\sqrt{1-\tau^{-1}}}\right) - i \pi\right]^2, \quad (\tau > 1). \eaa\r.
\eea
In the limit of very massive fermions, we have $A_{1/2}(\tau\ra 0)\ra 1$, so the contribution of the new heavy fermion fields to the $Hgg$ coupling will obey
\bea
\delta g_{Hgg}  \propto \sum_{M_i> m_H} \frac{Y_{ii}}{M_i},
\eea
where the sum is performed only over states that are more massive than the Higgs.
We can rewrite this sum as
\bea
\label{det}
\sum_i \frac{Y_{ii}}{M_{i}}-\sum_{M_i< m_H}\frac{Y_{ii}}{M_i} &=&  {\rm tr}(Y M^{-1})-\sum_{M_i< m_H}\frac{Y_{ii}}{M_i} \nonumber\\
&=& \frac{\d \log(\det M)}{\d v}-\sum_{M_i< m_H}\frac{Y_{ii}}{M_i}.
\eea 
Using the expression in the second line  proves to be very efficient for calculating this coupling, as one avoids having to explicitly compute the mass eigenstates.  

As an example of this calculation, we  consider a simple model with one vector-like doublet $Q$, and one vector-like up-type quark $U$.  The mass terms are given by
\bea
\Delta {\cal L}=M_Q \bar Q Q+M_U\bar U U+\left(y_1 H \, \bar Q_L U_R+y_2 H\, \bar Q_R U_L +{\rm h.c.}\right)
\eea 
For simplicity let us suppose $M_Q,M_u\gg m_H, yv$, and also assume that there are no mixing effects with light states.  Then we find that the modification of the gluon coupling will be proportional to the following:
\bea
\sum \frac{Y_{ii}}{M_{i}}=\frac{\d \log(\det M)}{\d v} \simeq \frac{-2y_1 y_2 v}{M_Q M_U}.
\eea
In the next sections we will apply this trick to understand the behavior  of the $Hgg$ coupling in  realistic composite models.

\section{Toy Examples}
\label{sec:Intro} \setcounter{equation}{0} \setcounter{footnote}{0}
\label{toyexample}
As a warmup exercise, we return to the model summarized in Table~\ref{1GenMatter}, where a single vector-like composite $\bf 5$ is introduced to mix with the elementary top quark.  
We will first analyze the modification of the gluon
coupling, and then study the modifications of the SM top Yukawa couplings.
\subsection{Top Quark and Its Custodians}
The following analysis will be simplified by making a field redefinition of the composite quarks.  Using $\xi$ as defined in Eq.~(\ref{eq:xi}), we simply take
\beq\label{eq:Qrescaling}
\Qc \to  \xi^\dagger \Qc.
\eeq
This transformation will generate Higgs derivative interactions, but they are not important for  single Higgs production\footnote{These derivative interactions will always be antisymmetric in the fermion fields (the coupling is proportional to the generators of $SO(5)$), thus they are irrelevant for  single Higgs production.}, so we will ignore them in what follows. 

In terms of the redefined fields, including the interaction terms between the composite scalars and fermions, we have
\bea\label{twosite}
\Delta {\cal L} = M_5 \bar \Qc_R  \Qc_L
+\lambda_q \bar q_L P_q \xi^\dagger \Qc_R 
+ \lambda_t \bar t_R  P_t \xi^\dagger \Qc_L
+Yf(\bar\Qc_R \Sigma_0) ( \Sigma_0^\dagger  \Qc_L) +{\rm h.c.}
\eea
where again the standard vacuum $\Sigma_0$ is defined in Eq.~(\ref{eq:Sigma0}).  From this we find the following mass matrix for the charge $2/3$ fields:
\bea
&M_t=\begin{pmatrix}
{ 0 & \frac{\lambda_q(\cos (v/f)+1)}{{2}} &\frac{ \lambda_q(\cos (v/f)-1)}{{2}} & \frac{ i\lambda_q \sin (v/f)}{\sqrt 2} \cr
 \frac{- i\lambda_t^*\sin (v/f)}{\sqrt 2} &  M_5 & 0 & 0 \cr
 \frac{- i\lambda_t^*\sin (v/f)}{\sqrt 2}  & 0 &  M_5 & 0 \cr
 \lambda_t^*\cos (v/f) & 0 & 0 &  M_5+ Y f}
 \end{pmatrix}.
\label{mtopccwz}
\eea
We make special note here of the fact that there is no Higgs dependence in the composite sub-block of the matrix, which is made particularly transparent after carrying out the field redefinition of Eq.~(\ref{eq:Qrescaling}).  Since the determinant is invariant under unitary transformations this property will be true in any basis.  Furthermore, since  $\xi$ commutes with the generator of electric charge, the composite part of the mass matrices will be independent of the Higgs for each different charge species individually.

Now for a light Higgs ($m_H\ll m_t$) following Eq.~(\ref{det}), we find that the top quark and its custodial partners will contribute to the Higgs coupling to gluons with a strength governed by
\bea\label{det1}
\frac{\d \log (\det M)}{\d  v} &=& \frac{2}{f} \cot \left( \frac{2v }{f} \right) \nonumber \\
&\simeq& \frac{1}{v}\l(1-\frac{4 v^2}{3f^2}\r).
\eea
Note that the value $v$ of the Higgs VEV is related to the electroweak symmetry breaking VEV, $v_{\rm SM} = (\sqrt 2 G_F)^{-1/2} = 246 \ {\rm GeV}$, by matching the expression for the $W$ mass, 
\bea
m_W &=& \frac{g f}{2} \sin \left( \frac{v}{f} \right) \nonumber \\
&=& \frac{g \, v_{\rm SM}}{2},
\eea
i.e.
\bea
v\simeq v_{\rm SM}\l( 1+\frac{1}{6}\frac{v_{\rm SM}^2}{f^2}\r).
\eea
The overall modification of the Higgs coupling to gluons, Eq.~(\ref{det1}), can be recast accordingly:
\bea
\frac{1}{v}\l(1-\frac{4 v^2}{3f^2}\r) \rightarrow \frac{1}{v_{\rm SM}}\frac{\l( 1-\frac{2 v_{\rm SM}^2}{f^2}\r)}{\sqrt{1-\frac{v_{\rm SM}^2}{f^2}}}\simeq\frac{1}{v_{\rm SM}}\l(1-\frac{3}{2}\frac{v^2_{\rm SM}}{f^2}\r).
\eea

Thus the coupling's modification follows a simple trigonometric scaling with no dependence on the masses of the custodians, as  reported in \cite{Falkowski:2007hz,Low:2010mr,Furlan:2011uq}.  This is an important point, and we will consider the question of its generality below.

We have thus far seen that the contribution of the top-like states to the gluon fusion coupling is given by a rescaling that is independent of the custodians' masses.  For the top Yukawa coupling, however, this turns out not to be the case.  We have the following expression for the top mass:
\bea
\label{mtopSM}
m_t=\frac{y_t f}{2}\sin \l(\frac{2v}{f}\r)\left[1+\mathcal O\l( (\lambda v/(f M_*))^2\r) \right],
\eea
where the  $\mathcal O (\lambda^2)$ corrections arise from  diagrams as shown in Fig.~2. The effects of the wavefunction renormalization are given by the following:
\begin{figure}[t]
\begin{center}
\includegraphics{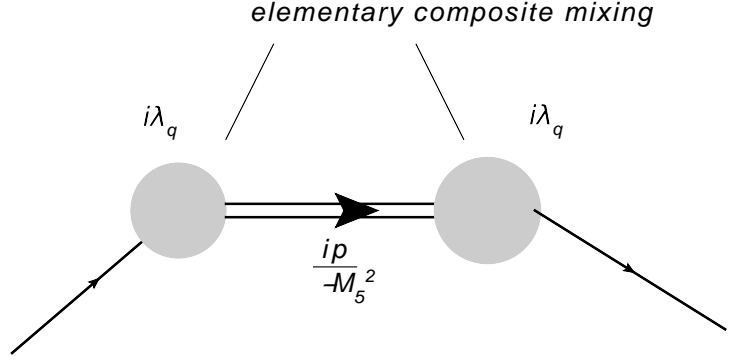}
\caption{Wavefunction renormalization from mixing with heavy states.}
\end{center}
\end{figure}
\bea
i\slashed{p}&\ra& i\slashed{p} Z_q\nonumber\\
Z_q&=&\l[1+\frac{|\lambda_q|^2}{2}\l(\frac{1+\cos^2(v/f)}{M_5^2} +\frac{\sin^2(v/f)}{(M_5+Y f)^2}\r)\r]\nonumber\\
Z_t&=&\l[1+|\lambda_t|^2\l(\frac{\sin(v/f)^2}{M_5^2}+\frac{\cos(v/f)^2}{(M_5+Y f)^2}\r) \r],
\eea
where we ignore  terms at higher order in $\lambda_{t,q}$. These effects lead to a modification of the expression of the top mass,
\bea
m &\ra& \frac{m}{\sqrt{Z_q}\sqrt{Z_t}} \nonumber \\
&= & \frac{y_t f}{2}\sin(2v/f)\l[1-\frac{|\lambda_t|^2}{2}\l(\frac{\sin^2(v/f)}{M_5^2}+\frac{\cos^2(v/f)}{(M_5+Y f)^2}\r) \right. \nonumber \\
&&\left. \hspace{3.1cm} -\frac{|\lambda_q|^2}{4}\l(\frac{1+\cos^2(v/f)}{M_5^2} +\frac{\sin^2(v/f)}{(M_5+Y f)^2}\r) \r],
\eea
and thus to the Yukawa coupling
\bea
\label{ytop}
y=\frac{\d m_t}{\d v} =\frac{2 m_t}{f \tan(2v/f)}+\frac{y_t f}{4}\sin(2v/f)\l(\frac{1}{M_5^2}-\frac{1}{(M_5+Y f)^2}\r)\l[\frac{\lambda_q^2}{2}-\lambda_t^2\r]
\eea
The first term of Eq.~(\ref{ytop}) is just a trigonometric scaling  coming from the nonlinearity of the Higgs boson, while the second term is related to wavefunction renormalization effects. We note  that  this wavefunction renormalization effect does not have fixed sign, i.e.  it can increase as well as decrease the SM fermion Yukawa coupling. Although the expansion in terms of the $\lambda/M_5$ becomes ill-defined in the limit of  full compositeness, the lesson to learn is that the Yukawa coupling of the top quark receives a significant correction which depends on the masses of the custodian fields, and that the sign of this correction is not fixed.  We plot the results of the numerical calculation in Fig.~\ref{figtoy5}, where we take $f=800$ GeV, $M_*=\sqrt{\lambda_q^2+M_5^2}=3.2$ TeV, and $Y=3$.   The scan is carried out fixing $M_{*}$ such that the custodian fields can be made light, and so that the composite top partner does not become infinitely heavy in the fully composite limit \cite{Pomarol:2008bh}, while $\lambda_t$ is fixed by matching the top mass. We can see that the effect becomes large in  the case of the fully composite $t_L$, when there is a light custodian $t'$.
\begin{figure}[htb]
\begin{center}
\includegraphics[scale=0.9]{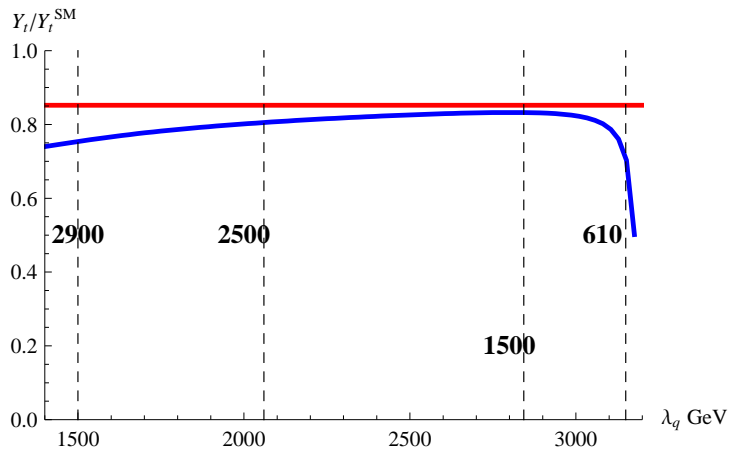}
\caption{
\label{figtoy5}
Modification of the top Yukawa coupling as a function of $\lambda_q$. The solid red line indicates the (constant) trigonometric rescaling for comparison, and the vertical dashed lines indicate the mass of the lightest $t'$ in GeV. Fully composite $t_L$ corresponds to  $\lambda_q\ra 3.2$ TeV,  $\sin\theta_q\ra 1$.}
\end{center}
\end{figure}

\subsection{Determinant Properties and Spurion Analysis}

As we have shown in the previous sections the coupling of the Higgs with gluons is controlled by the determinant of the fermion mass matrix. From   Eq. \ref{mtopccwz} we can immediately see that the only dependence of the determinant on the global symmetry breaking parameters $\lambda_{q,t}$ will enter as
\bea
\hbox{det} M_t\propto \lambda_q\lambda_t^*.
\eea
There are no terms with higher powers of $\lambda_{t,q}$, as is seen by noting that the global symmetry breaking parameters $\lambda_{q,t}$ appear only in the first row (column) of the mass matrix, Eq.~(\ref{mtopccwz}). Note that this property holds in any model based on partial compositeness where elementary fields get their masses only through linear mixing with composite states, and where the global symmetry of the composite sector is broken only by this mixing.
 
 We now consider the symmetry properties of our setup to provide a better understanding of the dependence of the modified SM couplings on the Higgs field.  We proceed by promoting the couplings $\lambda_q$ and $\lambda_t$ to spurions in order to make our Lagrangian, Eq.~(\ref{twosite}) formally $SO(5)$-invariant.   We thus take\footnote{We choose to split the electroweak doublet $\hat\lambda_q$ simply in order to allow separate analyses of the up- and down-type fields.}
\bea
\label{spur}
\bar q_L\lambda_qP_q&\ra& \bar q_L\hat\lambda_q =\bar t_L\hat\lambda_q^t+\bar b_L\hat\lambda_q^b, \nonumber \\
\bar t_R\lambda_tP_t &\ra& \bar t_R\hat\lambda_t.
\eea
Explicitly, this amounts to
\bea
\hat\lambda_q^t=\frac{\lambda_q}{\sqrt{2}}\l(\baa{c}0\\0\\i\\1\\0\eaa\r), \qquad
\hat\lambda_q^b=\frac{\lambda_q}{\sqrt{2}}\l(\baa{c}1\\-i\\0\\0\\0\eaa\r), \qquad
\hat\lambda_t\equiv \lambda_t\l(\baa{c}0\\0\\0\\0\\1\eaa\r),
\eea
where the $\hat\lambda_i$ transform under $U(1)_{\rm EM}$ of the elementary sector and the global $SO(5)$ according to
\bea
\hat \lambda_i \mapsto  e^{-i\alpha q_i}\cdot  \hat\lambda_i\cdot  {\cal G},
\eea
with ${\cal G} \in SO(5)$ and $q_i$ the corresponding electric charge. 
 The Lagrangian is formally invariant under  $SO(5)$ and $U(1)_{\rm EM}$ once these promotions are made, as will be the determinant of the mass matrices\footnote{One might expect this to apply only for the {\it total} mass matrix including all fields.  However, the mass matrices factorize, i.e. $\det M_{\rm tot}=\det M_t \det M_{-1/3} \det M_{5/3}$, where $\det M_{-1/3}$ is  the determinant of the composite sub-block of the matrix for the $b$-like quarks (similarly for $M_{5/3}$).  We have seen that these will be independent of $v$,   and thus they can be safely ignored here.}. 
 Thus we can deduce that in the case of the top-like fields, we must have
\bea
\det M_t\propto (\Sigma^\dagger \hat\lambda^t_q)(\hat\lambda_t^\dagger\Sigma)\propto 
\lambda_q\lambda_t^* \sin\l(\frac{2v}{f}\r),
\eea
as one can indeed check by direct computation.  We can see clearly now that  the dependence of the determinant on $v/f$ can be factorized, i.e.
\bea
\det M_t= (\Sigma^\dagger \hat\lambda^t_q)(\hat\lambda_t^\dagger\Sigma) \cdot  P(M_*,Y_*,f),
\label{detdep}
\eea
where $P(M_*,Y_*,f)$ is some polynomial  of the masses $M_*$ and Yukawa couplings $Y_*$ of the composite sector (i.e. $SO(5)$ singlet quantities), but which has no dependence on $v$. 
This is a crucial result, as the quantity
$\d_v \log (\det M_t)$ will not depend on $P(M_*,Y_*, f)$.
So we can see that in the limit 
$m_h\ll m_t$, corrections to the $Hgg$ coupling are described only in terms of the trigonometric rescaling, and there is no dependence on the actual masses of the composite fields. 

One might question whether this result will hold for the $5D$ models where we have Kaluza-Klein (KK) infinite tower of resonances, in this case the elementary fields will couple not to just one resonance but to the whole tower,
\bea
\lambda_q \bar q_L P_q \xi^\dagger \Qc_R \ra\sum_i
\lambda_q^{(i)}\bar q_L P_q \xi^\dagger \Qc_R^{(i)}.
\eea
However, since these resonances $\Qc^{(i)}$ have the same global $SO(5)$ quantum numbers we can always make the $\xi$-independent field redefinition
\bea
\tilde\Qc=\frac{\sum_{i}\Qc^{(i)}\lambda_q^{(i)}}{\sqrt{\sum_i\l(\lambda_q^{(i)}\r)^2}}, \quad
\tilde\lambda_q=\sqrt{\sum_i\l(\lambda_q^{(i)}\r)^2},
\eea
whereby one can see that the  elementary $q_L$ mixes only with $\tilde\Qc_L$ via the operator
\bea
\tilde\lambda_q \bar q_L P_q \xi^\dagger \tilde\Qc_R.
\eea
Thus, this case is similar to the one with the single resonance we considered before.

We can now consider the extent to which the preceding result (Eq.~(\ref{detdep})) is true in general.  Note that in order to derive  Eq.~(\ref{detdep}), we used only the fact  that the single $SO(5)$-invariant was $(\hat\Sigma^\dagger \lambda_q)(\hat\lambda_t^\dagger\Sigma)$, but generically this is not the case. For example, in the case where the left-handed elementary top  mixes with a ${\bf 5}$ as well as a ${\bf 10}$  from the composite sector, then we have two spurions $(\hat\lambda_q^{(5)},\hat\lambda_q^{(10)})$ from which we can form invariants. In this case, Eq.~(\ref{detdep}) is no longer true; one finds instead
\bea
\det M=
(\Sigma^\dagger \hat \lambda_q^{(10)} \lambda_t^\dagger)\cdot P_1(M_*,Y_*) +(\Sigma^\dagger \hat \lambda_q^{(5)} )( \hat\lambda_t^\dagger \Sigma) \cdot P_2(M_*,Y_*),
\eea
where
\bea
\Sigma^\dagger \hat \lambda_q^{(10)} \lambda_t^\dagger \propto \sin(v/f) \quad {\rm and} \quad
(\Sigma^\dagger \hat \lambda_q^{(5)} )( \hat\lambda_t^\dagger \Sigma)\propto \sin(2v/f).
\eea
In this case there will be some explicit composite mass dependence in the $Hgg$ coupling, so that by choosing appropriate values of the composite parameters $(M_*,Y_*)$ we can enhance as well as reduce the overall coupling (we refer the reader interested in the explicit model  to  Appendix \ref{topmodel}, where we present one example).  This will be true also in the case where the  top mixes with  higher dimensional representations, e.g. the $\bf 14$ (symmetric traceless).
For example in the model
with ${\bf 14}$ and ${\bf 5}$, where  $t_L$ mixes with ${\bf 5}$ and $t_R$ with ${\bf 14}$  there are two invariants
 \bea
&\hat\lambda_t^{14}\ra {\bf 14},~\hat\lambda_q^{5}\ra {\bf 5}\nonumber\\
&(\Sigma\hat\lambda_t^{14}\Sigma)(\Sigma\hat\lambda_q^{5})\propto \sin(v/f)\l(1-5\cos^2(v/f)\r),\nonumber\\
&(\Sigma\hat\lambda_t^{14}\hat\lambda_q^{5})\propto\sin(v/f)
\eea
so there again will be some explicit composite mass dependence in the $Hgg$ coupling.  

To recap the preceding arguments, what we have seen in the simplest two-site setups is that the custodial partners of the top and top itself conspire to produce a gluon coupling to the Higgs that follows a simple trigonometric rescaling.  The top Yukawa coupling, on the other hand, is seen to receive a more complicated modification.  The important point here is that the gluon coupling has a form that is not simply dictated by the top Yukawa.

Before proceeding further, we comment about the effects of the exotic charge $5/3$ fields.  In particular, one might wonder about their contribution to the $Hgg$ coupling.  We point out that this calculation becomes trivial in the basis of Eq.~(\ref{twosite}), where all the Higgs interactions are moved to the mixing between elementary and composite sectors.  Then since the exotic fields do not mix with the elementary sector, we see that they cannot contribute to the coupling of the Higgs to gluons.

\subsection{Bottom Quark and Its Custodians}
We  now   turn the discussion to the bottom quark, where one confronts the notable difference in that the lightest mode in the spectrum no longer satisfies $m \gg m_h$.  The computation of the gluon coupling is thus modified and the overall behavior  turns out to be quite different, provided the SM $b$ quark contains a substantial composite element, which might be the case for the left-handed $b_L$ quark.  

We begin the discussion by identifying certain conditions that the bottom-type quarks must satisfy in order to produce a noticeable  effect. First, we repeat that the masses of the SM fields will have more complicated dependence on $v/f$ than the total determinant, as in Eq.~(\ref{mtopSM}):
\bea
m_{b,t}\propto \lambda_q \lambda_{b,t} \cdot F(v/f)\cdot \l(1+O(\lambda v/(fM_*))^2 \r),
\eea
while
\bea
\det M_{b,t}\propto \lambda_q\lambda_{b,t} \cdot F(v/f).
\eea
Here $F(v/f)$ is some trigonometric function satisfying $F(0)=0$, as it should since one finds massless modes in the absence of EWSB.
In order to evaluate the $b'$ contribution to the $Hgg$ coupling we again need to perform the sum Eq.~(\ref{det}) over the heavy fields.  Unlike the top quark, however, the SM model $b$ quark must be excluded from the sum.  We modify the expression accordingly:
\bea
\sum_{M_i>m_H} \frac{Y}{M_i}&=&{\rm tr} \l(\hat Y_b M_b^{-1}\r)-\frac{y_b}{m_b}\nonumber \\
&=& \frac{\d \log (\det M_b)}{\d v}-\frac{y_b}{m_b},
\label{bcon}
\eea
where $\hat{Y_b}$ is the matrix of the Yukawa couplings of the $b$-like quarks and $y_b$ and $m_b$ are the Yukawa coupling and mass of the SM bottom.
We can rewrite this as
\beq
\sum_{M_i>m_H} \frac{Y}{M_i}=\frac{F'(v/f)}{fF(v/f)}-\frac{y_b}{m_b}.
\eeq
Finally, noting that
\bea
\frac{y_b}{m_b}&=&\frac{1}{m_b} \frac{\d m_b}{\d v}\nonumber \\ 
&=& \frac{F'(v/f)}{fF(v/f)}+\frac{1}{v} \cdot \mathcal O   \l(\frac{\lambda_q^2 v^2}{f^2 M_*^2}\r),
\label{bottom}
\eea
we find
\bea
\sum_{M_i>m_H} \frac{Y}{M_i} = 
\frac{1}{v}\cdot \mathcal O\l(\frac{\lambda_q^2 v^2}{f^2 M_*^2}\r).
\label{bcon}
\eea
In the limit of composite $t_L$, the ratio $\lambda_q/M_*$ can be large, so this effect will be numerically comparable or even larger than the one coming from the top sector, Eq~(\ref{det1}).  This shows that in the composite Higgs models, the  contribution of the $b$-like fields is important  for the  overall value of the $Hgg$ coupling, and also that these corrections are closely related to the modification of the bottom Yukawa.

With this observation, we have reduced the problem of finding the effective gluon coupling to that  of finding the modification of the Yukawa coupling of the SM $b$ quark. 
In particular, we are interested in the
$\mathcal O (\lambda_q^2)$ wavefunction renormalization that can produce such a modification.
We focus now on the conditions that must be met for such an effect to be present.

Since we are interested in the  SM bottom quark Yukawa interactions,  all the effects of the heavy fields can be parametrized in terms of higher dimensional effective operators, e.g.
\bea
(\bar q_L^{SM}\slashed {\d}q_L^{SM} ) (\hat\lambda_q^\dagger\Sigma\hat\lambda_q), \quad
(\bar b_R^{SM}\slashed {\d}b_R^{SM} )(\hat\lambda_b^\dagger\Sigma \hat\lambda_b), \quad
 \bar q_L^{SM} b_R^{SM} (\hat\lambda_q^\dagger \Sigma \hat\lambda_b),
\eea
where the exact contraction between $\Sigma $ which contains the Higgs field and the spurions $\hat \lambda_q,\hat\lambda_b$ depends on the representations of $SO(5)$ one chooses to mix with the elementary states.  As argued above, one expects the case of composite left-handed states to produce the most significant effects, so we focus here on the operator $(\bar q_L^{SM}\slashed{\d}q_L^{SM} )(\hat\lambda_q^\dagger\Sigma\hat\lambda_q) $.  We consider first the case where  $b_L$ mixes with a ${\bf 5}$ and $b_R$ with a ${\bf 10}$ of $SO(5)$.  Generally the Lagrangian in this case takes the form
\bea
\Delta {\cal L}=\bar b_L^{SM}  b_R^{SM}  (\hat\lambda_q^{b\dagger} \hat\lambda_b \Sigma) 
+\bar b_L^{SM}\slashed{\d} b_L^{SM}  \left| (\Sigma^\dagger \hat\lambda_q^b )\right|^2+
\bar b_R^{SM}\slashed{\d}b_R^{SM}
(\Sigma^\dagger\hat\lambda_b^\dagger\hat\lambda_b\Sigma ),
\eea
with coefficients suppressed for notational clarity.  There are also terms proportional to the $(\Sigma^\dagger\Sigma)$, but since they are Higgs-independent we can safely ignore them. We note though that
\bea
\hat\lambda_q^b\Sigma=0,
\eea
so there will be no corrections proportional to $\lambda_q^2$.  In the case where $b_L$ is mixed with the ${\bf 10}$, on the other hand,
one has an additional operator
\bea
\bar b_L^{SM}\slashed{\d}b_L^{SM} (\Sigma^\dagger \hat\lambda_q^{b \dagger} \hat\lambda_q^b \Sigma)
\eea
which is generically nonzero.   In the models where $b_L$ mixes with a ${\bf 10}$, we conclude that one  can realize  a  modification of the Yukawa coupling to the SM bottom quark.  We now demonstrate this explicitly. 

\subsection{The Single {\bf 10} Model: The Gluophilic $b$-Phobic Higgs}
\begin{figure}[thb]
\begin{center}
\includegraphics[scale=0.9]{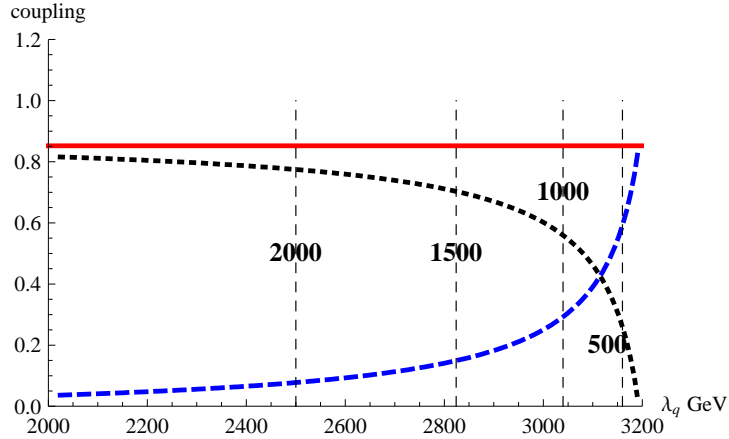}
\caption{Couplings in the model with a single composite  {\bf 10}.  The solid red line indicates the simple trigonometric rescaling for comparison.  The dotted  black line shows the bottom Yukawa relative to its SM value, and the dashed blue line shows the contribution of the bottom quark to the gluon coupling relative to the SM value in the large top mass limit.  Vertical dashed lines indicate the mass of the lightest composite $b'$ in GeV.
\label{single10} }
\end{center}
\end{figure}

The simplest model where we will have $b'$
effects will be the model with just one composite fermion multiplet in the antisymmetric $\bf 10$ of $SO(5)$, with
\bea
\Delta {\cal L}_{10}&=&M_{10}\, {\rm tr}(\bar{\Qc}_R \Qc_L)
+Y_*(\Sigma^\dagger \bar \Qc_R\Qc_L\Sigma)\nonumber\\
&+& \bar t^{SM}_R {\rm tr}(\hat\lambda_t^\dagger\Qc_L)+\bar b^{SM}_R {\rm tr}(\hat\lambda_b \Qc_L)+ {\rm tr}(\bar{\Qc}_R\hat\lambda_q)q_L^{SM},
\eea
from which one finds the mass matrix for the $b$ quarks:
\bea
M_b=
\begin{pmatrix}
{ 0 & \lambda _q & 0 & 0 \cr
 0 & \frac{1}{2} Y_* \cos ^2\left(\frac{v}{f}\right)+M_{10} & \frac{\cos \left(\frac{v}{f}\right) \sin \left(\frac{v}{f}\right) Y_*}{2 \sqrt{2}} &
   -\frac{\cos \left(\frac{v}{f}\right) \sin \left(\frac{v}{f}\right) Y_*}{2 \sqrt{2}} \cr
 \lambda _b & \frac{\cos \left(\frac{v}{f}\right) \sin \left(\frac{v}{f}\right) Y_*}{2 \sqrt{2}} & \frac{1}{4} Y_* \sin ^2\left(\frac{v}{f}\right)+M_{10} &
   -\frac{1}{4} \sin ^2\left(\frac{v}{f}\right) Y_* \cr
 0 & -\frac{\cos \left(\frac{v}{f}\right) \sin \left(\frac{v}{f}\right) Y_*}{2 \sqrt{2}} & -\frac{1}{4} \sin ^2\left(\frac{v}{f}\right) Y_* & \frac{1}{4}
   Y_* \sin ^2\left(\frac{v}{f}\right)+M_{10} 
   }
\end{pmatrix}.
\eea
From here, we can easily calculate the effects on the modification of the SM $b$ Yukawa coupling, as well as new contribution to the $Hgg$ coupling.
We proceed numerically, taking $f=800$ GeV, $M_{*}=3.2$ TeV, $Y_*=3$, and again holding  $\sqrt{\lambda_q^2+M_{10}^2}=M_*$ fixed.  Results are shown in Fig.~\ref{single10}.  One can see that these two effects are inversely related, as anticipated in Eq.~\ref{bcon}, and that the effects are maximal in the limit of a fully composite $b_L$. This large modification of the $Hgg$ coupling comes from the diagrams with light composite $b'$ circulating in the loops and effect becomes large in the limit when the mass of the $b'$ becomes light. 
 
In the limit of  full $b_L$ compositeness, we find that the Higgs becomes $b$-phobic and gluophilic. This behavior can lower the LEP bound on the Higgs mass below $114$ GeV, while at the same time providing an increased production cross section through gluon fusion at the LHC. We will discuss the LEP bounds and the latest LHC constraints in Section~\ref{LEP}.
 
We would like to comment that the model with a single ${\bf 10}$ can easily be made compatible with 5D holographic models: one has only to enlarge  the composite sector to include three ${\bf 10}$ multiplets instead of one.  In this case each elementary field will mix with a separate composite multiplet. 

\section{5D-inspired $({\bf 10,10,5})$ Model}
\label{sec:Intro} \setcounter{equation}{0} \setcounter{footnote}{0}
In the previous section we saw that when $b_L$ is mixed with a ${\bf 10}$, a large modification of both the bottom Yukawa and the gluon couplings arises due to the custodian $b'$. In this section we  present a 5D-inspired model based on the composite fermion content $\Tc^{5},\Qc^{10},\Bc^{10}$, with superscripts indicating the representation for each field. In this model, each SM field  will mix 
with a separate composite multiplet, as is realized in holographic models.
 The Lagrangian of the model will be given by   the following:
\bea
&\Delta {\cal L}_{10}=M_{10} \, {\rm tr} \l(\bar \Bc_R^{10} \Bc^{10}_L\r)+M_5^t(\bar{\Tc^5_R}\Tc^5_L)+M_{10}^q \l(\bar \Qc_R^{10} \Qc^{10}_L\r)+\nonumber\\
&Y_b^{(1)}f \Sigma^\dagger \bar \Qc_R^{10} \Bc_L^{10}\Sigma+
Y_b^{(2)} f \Sigma^\dagger \bar \Qc_L^{10} \Bc_R^{10}\Sigma+Y_t^{(1)} f\Sigma^\dagger \bar \Qc_R^{10} \Tc^5_L+Y_t^{(2)} f\bar \Tc^5_R \Qc^{10}_L\Sigma+\tilde Y f \Sigma^\dagger \bar \Qc_R^{10} \Qc_L^{10} \Sigma \nonumber\\
&+\bar t^{SM}_R (\hat\lambda_t^\dagger \Tc^5_L)+\bar b^{SM}_R \, {\rm tr}\l(\hat\lambda_b \Bc^{10}_L\r)+ {\rm tr}(\bar \Qc^{10}_R\hat\lambda_q)q_L^{SM}
\eea
The couplings $\tilde Y$ are not necessary to reproduce the masses of the SM fields, but they are allowed by the symmetries and thus we include them for completeness. The discussion of the $t$-like and $b$-like fields proceeds in the same way as for the model with single ${\bf 5}$  or ${\bf 10}$. However, the overall dependence of the determinant of the top fields will differ since $t_L^{SM} $ now mixes with a ${\bf 10}$ while $t_R^{SM}$ mixes with a ${\bf 5}$.  Following
 Eq.~(\ref{detdep}), we find
\bea
\det M_t\propto (\Sigma^\dagger \hat\lambda_q^t\hat\lambda_t)\propto\sin\l(\frac{v}{f}\r),
\eea
where $\hat\lambda_{q},\lambda_t$ transform as a ${\bf 10}$ and ${\bf 5}$, respectively.
Note that the argument of the sine has changed by a factor of two, giving a modified contribution to the gluon coupling relative to the models involving mixing with a single $\bf 5$.

\begin{figure}
\begin{center}
\includegraphics[scale=0.9]{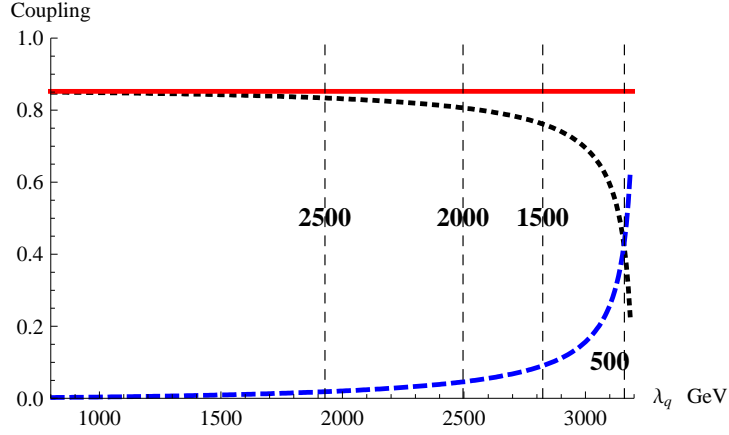}
\caption{Couplings for the ${\bf 10,10,5}$ model, with curves as in Fig.~\ref{single10}.\label{10105}}
\end{center}
\end{figure}

\begin{figure}
\begin{center}
\includegraphics[scale=0.8]{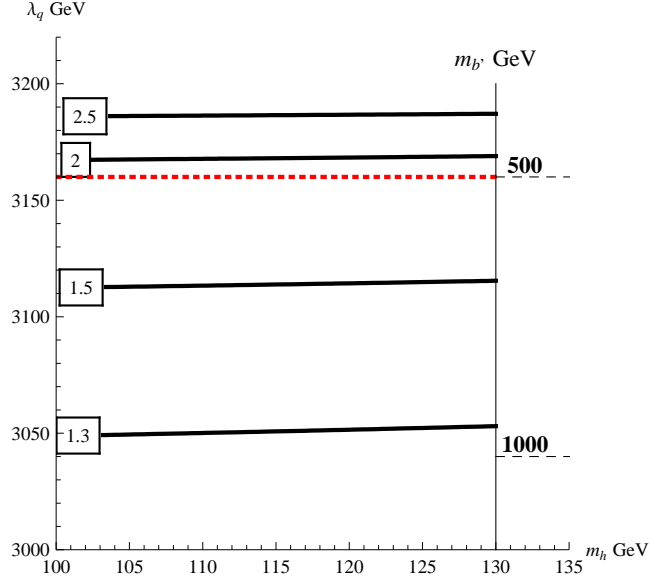}
\caption{Contour plot for the rescaling of the $\sigma(gg\ra H)$ \label{hgglambdaq} in the $(m_h,\lambda_q)$ plane.  The red dotted line indicates the constraint from $V_{tb}$.  We indicate the corresponding values of the $b'$ mass on the right side of the plot.}
\end{center}
\end{figure}

We can again proceed numerically, with results  presented in Figs.~\ref{10105} and \ref{hgglambdaq}.  Fig.~\ref{10105}  shows the dependence of the bottom quark Yukawa and gluon couplings  on the compositeness ($\lambda_q$) of the left-handed $b$ quark.  For the computation, we have set
$f=800$ GeV, $Y_i=3$, $M_*=4f=3.2$ TeV, varying $\lambda_q$ while keeping $M_*=\sqrt{\lambda_q^2+(M_{10}^q)^2}$ fixed.
As expected, for the small values of  $\lambda_q$
the new  effects are dominated by the trigonometric rescaling of the Yukawa coupling $m_b\propto\sin (2v/f)$.   Towards  larger values of $\lambda_q$,  however, wavefunction effects start to dominate, and we again find that the Higgs becomes gluophilic and $b$-phobic. Fig. \ref{hgglambdaq} shows overall modification of the production cross section from gluon fusion including the loops with SM top, $t'$ and $b'$.

\subsection{$H\gamma\gamma$}
For a light Higgs, $H\ra \gamma\gamma$  provides one of the most promising discovery channels.
We must therefore consider the coupling (and branching fraction) of $H \to \gamma\gamma$ for our specific model to determine its discovery prospects. The photon coupling in the SM is produced dominantly by one-loop diagrams involving circulating $W$, $Z$, and $t$ fields. In composite models, we thus need to account for loops with new bosonic and fermionic fields, as well as the modifications induced in the tree-level SM couplings. The effects coming from new fermions are similar to those contributing to $Hgg$, and the calculation proceeds in exactly the same way. The contribution of the new bosonic degrees of freedom 
 can also be carried out numerically using known formulas for the loop integrals (cf. for example \cite{Ellis:1975ap,Gunion:1989we}).   For simplicity, we consider   here only the limit where the mass of the composite vector bosons is much larger than the PNGB decay constant, $f$. In this limit, the dominant effects come from the nonlinearity of the Higgs boson, where the mass of the $W$ is given by
\bea
M_W^2=\frac{g^2 f^2}{4}\sin^2 (v/f), 
\eea
implying a shift in the coupling $g_{HWW}$:
\bea
\label{gammaw}
\frac{g_{HWW}}{g_{HWW}^{SM}}=\cos \l(\frac{v}{f}\r)=\sqrt{1-\frac{v_{SM}^2}{f^2}}
\eea
Accordingly,  we can simply scale the SM $W$ boson contribution to the $H\gamma\gamma$ coupling using this rescaling.
\begin{figure}[htb]
\begin{center}
\includegraphics[scale=0.8]{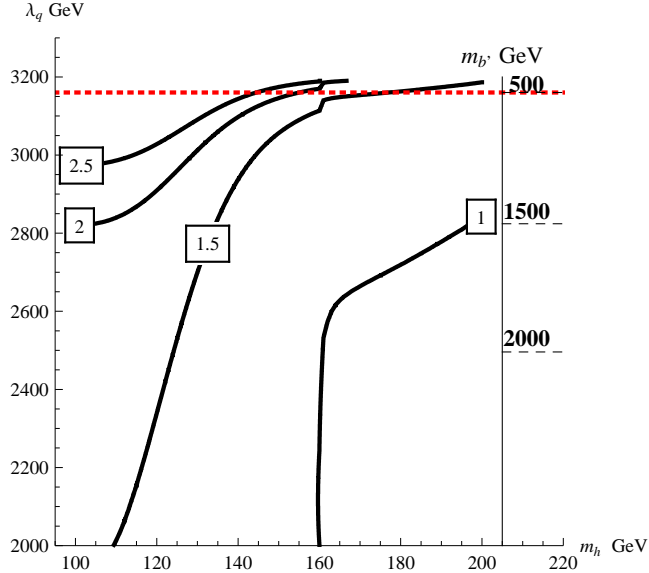}
\caption{
\label{sigmabr}
Contours of $\sigma (gg\ra H)\times {\rm BR}(H\ra \gamma\gamma)$ in the $(\bf 10,10,5)$ model, relative to the SM prediction. 
The red dashed horizontal line indicates the maximal value of $\lambda_q$ allowed by the $V_{tb}$ constraint. 
On the right side of the plot, we  show corresponding masses of the $b'$. We find an enhanced signal due to reduction of the $Hbb$ coupling, and to an increase of the $Hgg$ coupling.}
\end{center}
\end{figure}
With this, we plot contours of the rescaled $(gg\ra H\ra \gamma\gamma)$ signal in  Fig.~\ref{sigmabr}.
We can see that in the composite $q_L$  limit, the signal can be enhanced by up to a factor of three; this comes from the  suppression of the bottom Yukawa coupling as well as an enhancement of the gluon fusion cross section. 

\subsection{Constraints from LEP, Tevatron, and LHC}
\label{LEP}
We discuss now the known constraints for the models we have explored.   There are precision measurements regarding the compositeness of the SM fermions, and direct Higgs searches that constrain the available parameter space.

First, we note that generically the $Z\bar bb$ presents a very strong constraint on scenarios involving composite third generation quarks.  In our case, though, there in an enhanced  custodial symmetry \cite{Agashe:2003zs,Agashe:2006at,Contino:2006qr} suppressing new contributions.  Even in the limit of composite $b_L$, we find 
$\delta g_{Z\bar bb}\lesssim 0.001 \times g_{Z\bar bb}^{\rm SM}$,\footnote{The coupling $g_{Z\bar bb}$ as well as $V_{tb}$ was calculated numerically for $2.4$ TeV composite vector bosons.}  satisfying  the current experimental bounds. 

There is also an important bound on the CKM element $|V_{tb}|\gtrsim 0.77$ \cite{Group:2009qk}. We have checked numerically that this bound is saturated  for  $\lambda_q\sim 3160, \ m_{b'}\sim 470 $ GeV given our choice of  parameters.  The results for the overall modification of the gluon fusion production cross section are presented on Fig \ref{hgglambdaq}.  The plot clearly illustrates that for the given set of parameters we can easily enhance the total production cross section by a  factor of $\approx 1.8$ and still be consistent with the current bound on $V_{tb}$.

\begin{figure}[htb]
\begin{center}
\includegraphics[scale=0.9]{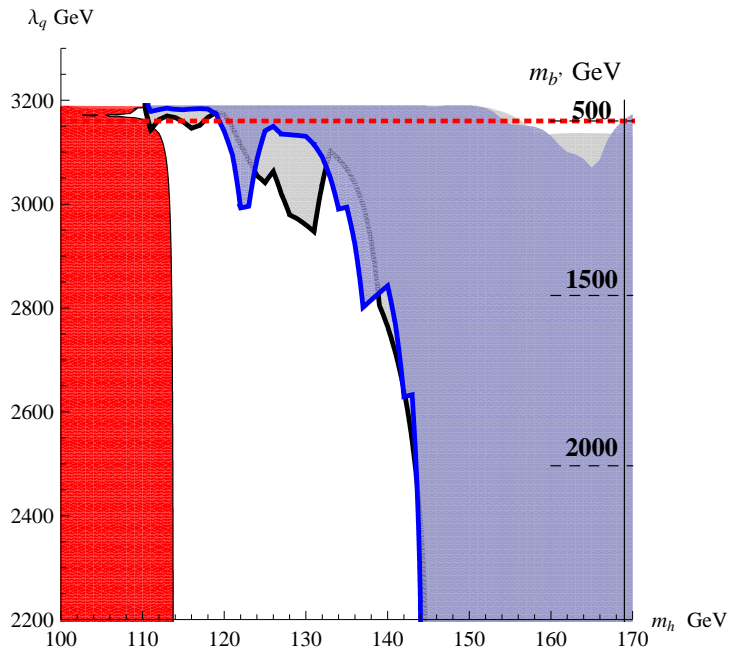}
\caption{
\label{constrlep}
Constraints from LEP and LHC in the ($m_H,\lambda_q$) plane. The red region is ruled out by LEP, while the grey (blue) area is constrained by CMS (ATLAS).  The dashed red line indicates the upper bound on compositeness as constrained by measurements of $V_{tb}$. On the right hand side, we have shown the masses of the $b'$ for given $\lambda_q$.}
\end{center}
\end{figure}

We now consider the constraints on the model from  LEP and recent results from the LHC. 
The LEP bounds can be easily checked using the code from the HiggsBounds group \cite{arXiv:0811.4169,arXiv:1102.1898}; we show these results in Fig.~\ref{constrlep}. The red shaded area corresponds to the region in the ($\lambda_q,m_h$) plane which is excluded by LEP. 
We can see that  increasing degree of compositeness of $b_L$ relaxes the  LEP bounds on the Higgs mass, as $H\ra \bar{b}b$ is suppressed, allowing the Higgs to be as light as 108 GeV.
 
Finally we comment on the recent bounds coming from the LHC. Since the couplings of the Higgs field in our model are not the same for the different fields we cannot simply rescale the combined exclusion plots from ATLAS \cite{ATLAS} and CMS \cite{CMS-PAS-HIG-11-022}.
However, in order to gain an idea of the constraints on $b_L$ compositeness, we can carry out the following exercise: for the every channel of the Higgs search  presented in \cite{CMS-PAS-HIG-11-022,ATLAS}, we calculate the rescaling of the signal due to the modifications of the Higgs couplings and then check whether the point is excluded by each channel independently.
These results are presented in Fig.~\ref{constrlep}: we can see  that the Higgs with mass above 140 GeV is nearly ruled out,
while the case of a light Higgs and a composite $b_L$ is allowed. We would like to emphasize, though,  that  the goal of this exercise was simply to gain an approximate idea of the allowed parameter space.  We cannot substitute this simplified rescaling for a full analysis, which is far beyond the scope of this paper.

\section{Conclusion}
\label{sec:Intro} \setcounter{equation}{0} \setcounter{footnote}{0}
We would like to conclude by restating the main results of this work. We have presented a detailed  analysis of the $Hgg$ coupling in models where the Higgs is a composite PNGB field. We investigated generic properties of this coupling and its dependencies on the parameters of the model. 
We have identified a class of  models where the contribution  of only the charge $2/3$ fields results  in a suppressed value of the $Hgg$ coupling  compared to its SM value.  We have also shown that this suppression (for the light Higgs) does not depend on the masses of the $t'$,  confirming results  known previously for specific cases \cite{Falkowski:2007hz,Low:2010mr,Furlan:2011uq}.
We constructed a model where the contribution of $2/3$ fields enhances $Hgg$ coupling,  and further we saw that this type of models is potentially dangerous due to Higgs-mediated flavor violation. 
Interestingly, we have found
that the modifications of the $Hgg$ and $Ht\bar {t}$ couplings turn out to be quite independent quantities, and that modification of the top Yukawa coupling depends strongly on the masses of the composite $t'$.  

Finally, we studied the effects of the composite $b'$ on the $Hgg$ coupling. We have shown that in the fully composite $b_L$ limit, effects of the composite $b'$ are important and can lead to an overall enhancement of the $Hgg$ coupling compared to the SM value. We discuss the phenomenology of the composite $b_L$ limit and point out an interesting $b-$phobic gluophilic Higgs limit. Within this limiting case, we investigate the modification of the Higgs production and subsequent decay to a two photon final state. 
We have compared the parameter space of the model to constraints coming from past and current collider experiments, showing that a sizable subspace remains viable.  Within the surviving parameter space, one finds that concrete predictions can be made that would have a significant bearing on forthcoming results from the LHC.

\section*{Acknowledgments}
We would like to thank R. Contino for  stimulating discussions, encouragement, and comments on the manuscript. We also thank N. Vignaroli for collaboration during initial stages of the work, and K. Agashe for comments. We are grateful  to A. Falkowski for pointing out the importance of the constraint on $V_{tb}$.

\appendix

\section{Generators and Representations of $SO(5)$}
\label{generators} \setcounter{equation}{0} \setcounter{footnote}{0}
In this section we review the group theory we've used throughout for the coset $SO(5)/SO(4)$.  The generators for  fundamentals of  $SO(5)$  are given by
\bea\label{eq:generators}
T^a_{L,ij}&=&-\frac{i}{2}
\l[\frac{1}{2}\epsilon^{abc}\l(\delta^b_i
\delta^c_j-\delta^b_j\delta^c_i \r)+\l(\delta^a_i\delta^4_j-\delta^a_j \delta^4_i \r)\r], \nonumber\\
T^a_{R,ij}&=&-\frac{i}{2}
\l[\frac{1}{2}\epsilon^{abc}\l(\delta^b_i
\delta^c_j-\delta^b_j\delta^c_i \r)-\l(\delta^a_i\delta^4_j-\delta^a_j \delta^4_i \r)\r], \nonumber\\
T^a_{C,ij}&=&-\frac{i}{\sqrt{2}}
\l[\delta^a_i\delta^5_j-\delta^a_j \delta^5_i \r].
\eea
Here $T_{L,R}$ denote respectively  the generators of the $SU(2)_{L,R}$ subgroups, and $T_C$ the coset generators. 

The decomposition of the fundamentals of $SO(5)$ can be determined in a straightforward way by expressing them  as sums of eigenvectors of $T^3_{L,R}$.
We have
\beq
{\bf 5} = \frac{1}{\sqrt 2} \begin{pmatrix}
{q_{++} + q_{--}  \cr
iq_{++} -i q_{--} \cr
q_{+-} +q_{-+} \cr
iq_{+-}  - i q_{-+} \cr
\sqrt 2 q}
\end{pmatrix},
\eeq
with each subscript denoting the eigenvalue of $T^3_L$ and $T^3_R$ respectively (hypercharge is identified with the latter). We see then that the $\bf 5$ consists of a bidoublet and a singlet of $SU(2)_L \times SU(2)_R$:
\bea
(2,2) & = & \begin{pmatrix} {q_{+-} & q_{++} \cr
q_{--} & q_{-+} }
\end{pmatrix}; \nonumber \\
1 &=& q.
\eea
Similarly we can decompose the antisymmetric ($\bf 10$) representation under the EW group.  We have
\bea
{\bf 10} = \frac{1}{2}\, \times \hspace{14.5cm} & \nonumber \\
\begin{pmatrix} 
{
 0 & u+u_1 & \frac{i (d-\chi )}{\sqrt{2}}+\frac{i (d_1-\chi_1)}{\sqrt{2}} & \frac{d+\chi }{\sqrt{2}}-\frac{d_1+\chi_1}{\sqrt{2}} & d_4+\chi_4 \cr
 -u-u_1 & 0 & \frac{d+\chi }{\sqrt{2}}+\frac{d_1+\chi_1}{\sqrt{2}} & \frac{i (d_1-\chi_1)}{\sqrt{2}}-\frac{i (d-\chi
   )}{\sqrt{2}} & -i (d_4-\chi_4) \cr
 -\frac{i (d-\chi )}{\sqrt{2}}-\frac{i (d_1-\chi_1)}{\sqrt{2}} & -\frac{d+\chi }{\sqrt{2}}-\frac{d_1+\chi_1}{\sqrt{2}} & 0 &
   -i (u-u_1) & t_4+T_4 \cr
 \frac{d_1+\chi_1}{\sqrt{2}}-\frac{d+\chi }{\sqrt{2}} & \frac{i (d-\chi )}{\sqrt{2}}-\frac{i (d_1-\chi_1)}{\sqrt{2}} & i
   (u-u_1) & 0 & -i (t_4-T_4) \cr
 -d_4-\chi_4 & i ({d_4}-{\chi_4}) & -{t_4}-{T_4} & i ({t_4}-{T_4}) & 0
}
\end{pmatrix}. \nonumber
\eea
\beq
\eeq
So we find $\bf 10={\bf (2,2)}+{\bf (1,3)}+{\bf (3,1)}$, where
\bea
({\bf 3,1})&=&(\chi,u,d), \nonumber \\
({\bf 1,3})&=&(\chi_1,u_1,d_1), \nonumber\\
{\bf (2,2)}&=&\begin{pmatrix}
{
\chi_4 & t_4\cr
T_4 & b_4
}
\end{pmatrix}.
\eea
Here $\chi$ stands for the exotic field with charge $5/3$, $t,t',u,u_1$ are the fields with charge $2/3$, and $d,d_1,b$ are fields with   charge $-1/3$.

\section{Non-minimal $t$-composite mixing }
\label{topmodel} \setcounter{equation}{0}
In this section we present a model based on $t_L$ mixing with both a ${\bf 5}$ and $\bf 10$ of $SO(5)$\footnote{A Similar model, where  $t_R$ mixes non-minimally with the composite sector can be constructed.}. We show that in this model, the $Hgg$ coupling coming only from $t$-like fields can be enhanced as well as reduced.

The model is described by the Lagrangian
\bea
\Delta {\cal L}_{10+5}&=&M_{10}\, {\rm tr}(\bar{\Qc}_R \Qc_L)
+Y_{10}(\Sigma^\dagger \bar \Qc_R\Qc_L\Sigma)\nonumber\\
&+& \bar t^{SM}_R {\rm tr}(\hat\lambda_t^\dagger\Qc_L)+\bar b^{SM}_R {\rm tr}(\hat\lambda_b \Qc_L)+ {\rm tr}(\bar{\Qc}_R\hat\lambda_q^{(10)})q_L^{SM}\nonumber\\
&+&M_5 \bar \Tc_R  \Tc_L
 +  (\bar\Tc_R \hat\lambda_q^{(5)})q_L^{SM}
+Y_5(\bar \Tc_R\Qc_L\Sigma)+\tilde Y_5 (\Sigma^\dagger \bar \Qc_R \Tc_L),
\eea
where $\Tc$ and $\Qc$ belong to the $\bf{5}$ and $\bf{10}$ representations, respectively and $t_R^{SM}$ mixes  only with  $\Qc$.  For the determinant of the mass matrix for the charge $2/3$ fields, we find
\bea
\det M_{t}\propto \sin \left(\frac{v}{f}\right) \left(\sqrt{2} f M_{10} \tilde Y_5 \lambda _{q}^{(5)}-\cos \left(\frac{v}{f}\right) \left(f^2 Y_5\tilde Y_5 +f M_5 Y_{10}\right) \lambda_{q}^{(10)}\right) \lambda_t.
\eea
In the limit where $\lambda_q^{(5)}$ (or $\lambda_q^{(10)})$ vanishes, we are back to the simple trigonometric rescaling $\det M_t\propto \sin(2v/f)$ or $\sin (v/f)$, respectively.
Expanding in powers of $v/f$,
\bea
\det M_t &\propto& \frac{v}{f}\l[1+\l(\frac{4 f^2 \lambda _{q}^{10} Y_5\tilde Y_5-\sqrt{2} f M_{10} \lambda _{q}^{(5)} \tilde Y_5+4 f M_5 Y_{10} \lambda _{q}^{(10)}}{-f^2 \lambda_{q}^{(10)} Y_5 \tilde Y_5+\sqrt{2} f M_{10} \lambda_{q}^{(5)} \tilde Y_5-f M_5 Y_{10} \lambda_{q}^{(10)}}\r)\frac{v^2}{6 f^2} \r]\nonumber\\
\frac{\d \log(\det M_t)}{\d v} &\simeq& \frac{1}{v}\l[1+\l(\frac{4 f^2 \lambda _{q}^{10} Y_5\tilde Y_5-\sqrt{2} f M_{10} \lambda _{q}^{(5)} \tilde Y_5+4 f M_5 Y_{10} \lambda _{q}^{(10)}}{-f^2 \lambda_{q}^{(10)} Y_5 \tilde Y_5+\sqrt{2} f M_{10} \lambda_{q}^{(5)} \tilde Y_5-f M_5 Y_{10} \lambda_{q}^{(10)}}\r)\frac{v^2}{3 f^2} \r]\nonumber\\
\frac{\d \log(\det M_t)}{\d v} &\simeq& \frac{1}{v_{\rm SM}}\l[1+\l(\frac{9 f^2 \lambda _{q}^{10} Y_5\tilde Y_5-3\sqrt{2} f M_{10} \lambda _{q}^{(5)} \tilde Y_5+ 9f M_5 Y_{10} \lambda _{q}^{(10)}}{-f^2 \lambda_{q}^{(10)} Y_5 \tilde Y_5+\sqrt{2} f M_{10} \lambda_{q}^{(5)} \tilde Y_5-f M_5 Y_{10} \lambda_{q}^{(10)}}\r)\frac{v^2_{\rm SM}}{6 f^2} \r]\nonumber\\
\eea
The second term in brackets controls the modification of the $Hgg$ coupling: the coupling is enhanced when this term is positive, and reduced when it is negative.  For example, if $\tilde Y_5 f=Y_5f=Y_{10} f=M_5=M_{10}$, then
\bea
\frac{\log(\det M_t)}{\d v} \propto 
\frac{1}{v_{\rm SM}}\l[1+\l(\frac{6\lambda_q^{(10)}-\sqrt{2}\lambda_q^{(5)}}{\sqrt{2}\lambda_q^{(5)}-2\lambda_q^{(10)}}\r)
\frac{v^2_{\rm SM}}{2 f^2} \r],
\eea
so that we have an overall enhancement of the $Hgg$ coupling if $\sqrt{2}\lambda_q^{(10)}<\lambda_q^{(5)}<3\sqrt{2}\lambda_q^{(10)}$. Note that this effect is not dependent on whether $t_L$ was fully composite or not, contrary to the $b'$ contribution discussed in the text.

Finally, we note that although the models with nonminimal elementary composite mixing can lead to very interesting phenomenology, they are potentially dangerous because of Higgs-mediated flavor violation  when such non-minimal mixing is present in the light quark sector \cite{Agashe:2009di}.

\pagebreak

\bibliographystyle{h-physrev}
\bibliography{lit}

\end{document}